\documentstyle[twocolumn,prc,aps,epsf]{revtex}

\preprint{NUC-MINN-02/3-T}

\newcommand{\be}{\begin{equation}}
\newcommand{\ee}{\end{equation}}

\newcommand{\ba}{\begin{eqnarray}}
\newcommand{\ea}{\end{eqnarray}}
\def\thalf{{\textstyle{\frac{1}{2}}}}
\def\oneth{{\textstyle{\frac{1}{3}}}}

\begin{document}
\draft

\title{Thermal conductivity of dense quark matter and 
cooling of stars}

\author{Igor A.~Shovkovy$^{*}$ and Paul J. Ellis} 

\address{School of Physics and Astronomy, University of Minnesota,
        Minneapolis, MN 55455, USA}

\date{June 3, 2002}
\maketitle

\begin{abstract} 

The thermal conductivity of the color-flavor locked phase of dense quark
matter is calculated. The dominant contribution to the conductivity comes
from photons and Nambu-Goldstone bosons associated with breaking of baryon
number which are trapped in the quark core. 
Because of their very large mean free path 
the conductivity is also very large. The cooling of the quark core arises
mostly from the heat flux across the surface of direct contact with 
the nuclear matter. As the thermal conductivity of the neighboring
layer is also high, the whole interior of the star should be nearly 
isothermal. Our results imply that the cooling time of compact stars with 
color-flavor locked quark cores is similar to that 
of ordinary neutron stars. 

\end{abstract}



\section{Introduction}

At sufficiently high baryon density the nucleons in nuclear matter should
melt into quarks so that the system becomes a quark liquid. It should be
weakly interacting due to asymptotic freedom \cite{ColPer}, however, it
cannot be described as a simple Fermi liquid. This is due to the
nonvanishing attractive interaction in the color antitriplet quark-quark
channel, provided by one-gluon exchange, which renders a highly degenerate
Fermi surface unstable with respect to Cooper pairing.  As a result the
true ground state of dense quark matter is, in fact, a color
superconductor \cite{old}.

Recent phenomenological \cite{W1S1} and microscopic studies
\cite{PR1-Son,2nd-wave,3rd-wave,us2} have confirmed that quark matter at a
sufficiently high density undergoes a phase transition into a color
superconducting state. Phenomenological studies are expected to be
appropriate to intermediate baryon densities, while microscopic approaches
are strictly applicable at asymptotic densities where perturbation theory 
can be used. It is remarkable that
both approaches concur that the superconducting order parameter (which
determines the gap $\Delta$ in the quark spectrum) lies between $10$ 
and $100$ MeV
for baryon densities existing in the cores of compact stars.  For a review
of color superconducting quark matter see Ref.~\cite{RW-Al-rev}.

At realistic baryon densities only the three lightest quarks can
participate in the pairing dynamics. The fact that the current mass of the
strange quark is more than an order of magnitude larger than the masses of
the up and down quarks suggests two limiting cases. The first extreme
limit assumes that the strange quark is sufficiently massive that it does
not participate in the pairing dynamics at all.  The corresponding quark
phase is the color superconducting phase involving two quark flavors
(2SC). Since it was suggested in Ref.~\cite{alf} that such a phase is
absent in compact stars, we focus on the second extreme case. This applies
when all three quark flavors have a mass which is small in comparison to
the chemical potential so that they
participate equally in the color condensation. The ground state is then
the so-called color flavor locked (CFL) phase \cite{ARW}. Here the
original gauge symmetry $SU(3)_{c}$ and the global chiral symmetry
$SU(3)_{L} \times SU(3)_{R}$ break down to a global diagonal ``locked"
$SU(3)_{c+L+R}$ subgroup. Because of the Higgs mechanism the gluons become
massive and decouple from the infrared dynamics. The quarks also decouple
because large gaps develop in their energy spectra. The breaking of the
chiral symmetry leads to the appearance of an octet of
pseudo-Nambu-Goldstone (NG) bosons ($\pi^{0}$, $\pi^{\pm}$, $K^{\pm}$,
$K^{0}$, $\bar{K}^{0}$, $\eta$). In addition an extra NG boson $\phi$ and
a pseudo-NG boson $\eta^{\prime}$ appear in the low energy spectrum as a
result of the breaking of global baryon number symmetry and approximate
$U(1)_{A}$ symmetry, respectively.

The low energy action for the NG bosons in the limit of asymptotically
large densities was derived in Refs.~\cite{CasGat,SonSt}. By making use of
an auxiliary ``gauge" symmetry, it was suggested in Ref.~\cite{BS} that
the low energy action of Refs.~\cite{CasGat,SonSt} should be modified by
adding a time-like covariant derivative to the action of the composite
field. Under a favorable choice of parameters, the modified action
predicted kaon condensation in the CFL phase. Some unusual properties of
such a condensate were discussed in Refs.~\cite{MirSho,Son-s}.

While the general structure of the low energy action in the CFL phase can
be established by symmetry arguments alone \cite{CasGat}, the values of
the parameters in such an action can be rigorously derived only at
asymptotically large baryon densities \cite{SonSt,BS}. Thus, in the most
interesting case of intermediate densities existing in the cores of
compact stars, the details of the action are not well known. For the
purposes of the present paper, however, it suffices to know that there are
9 massive pseudo-NG bosons and one massless NG boson $\phi$ in the low
energy spectrum. If kaons condense \cite{BS} an additional NG boson 
should appear.\footnote{Strickly speaking the additional NG boson
in a phase with kaon condensation has a small, but nonzero, mass as a
result of weak interactions that break strangeness. The estimated value
of the mass is of order 50 keV \cite{Son-s}.} These NG bosons 
should be relevant for the kinetic properties of dense quark matter.

It has been found \cite{JPS} 
that neutrino and photon emission rates for the CFL phase are very small
so that they would be inefficient in cooling the core of a neutron star.
The purpose of the present investigation is to determine quantitatively 
the thermal conductivity of the CFL phase of dense quark matter in order
to see whether it can significantly impact the cooling rate.  We shall 
argue that the temperature of the CFL core,
as well as the neighboring neutron layer which is in contact with the
core, falls quickly due to the very high thermal conductivities on both
sides of the interface. In fact, to a good approximation, the interior of
the star is isothermal. A noticeable gradient of the temperature appears
only in a relatively thin surface layer of the star where a finite flux of
energy is carried outwards by photon diffusion.

The paper is organized as follows. In the next section, we derive the
microscopic expression for thermal conductivity of the (pseudo-) NG
bosons in the CFL phase. In Sec.~\ref{mfp} we estimate the mean free path
$\ell$ of the NG boson in the CFL phase. Knowledge of this quantity is
crucial to a reliable calculation of the thermal conductivity. The role
of photons in affecting the thermal conductivity and specific heat
of the quark core is discussed in Sec.~\ref{photon}. In
Sec.~\ref{cool} we discuss our findings and their implication for the
cooling dynamics of a compact star with a quark core in the CFL phase. Our
conclusions are summarized in Sec.~\ref{conclusion}. In the Appendix we
present some useful formulae.

\section{Thermal conductivity}
\label{thcond}

A detailed understanding of the cooling mechanism of a compact star with
a quark core is not complete without a study of thermal conductivity
effects in the color superconducting quark core. The conductivity, as well 
as the other kinetic properties of quark matter in
the CFL phase, is dominated by the low energy degrees of freedom. It is
clear then that at all temperatures of interest to us $T\ll \Delta$,
it is crucial to consider the contributions of the NG 
bosons. In addition, there may be an equally important contribution
due to photons; this is discussed in
Sec.~\ref{photon}. Note that, at such small temperatures, the gluon 
and the quark quasiparticles become
completely irrelevant. For example, a typical quark contribution to a
transport coefficient would be exponentially suppressed by the factor
$\exp(-\Delta/T)$. Therefore, in the rest of this section, we concentrate
exclusively on the contributions of the NG bosons.

Let us start from the general definition of the thermal conductivity as a
characteristic of a system which is forced out of equilibrium by a
temperature gradient. In response to such a gradient transport of heat
is induced. Formally this is described by the following relation:
\be
u_{i} = - \kappa \partial_{i} T,
\ee
where $u_{i}$ is the heat current, and $\kappa$ is the heat
conductivity. As is clear from this relation, the heat flow would persist
until a state of uniform temperature is reached. The
higher the conductivity, the shorter the time for this relaxation.

In the linear response approximation, the thermal conductivity is given in
terms of the heat current correlator by a Kubo-type formula.  We
derive the expression for the heat (energy) current carried by a single
(pseudo-) NG boson field $\varphi$. The corresponding Lagrangian density
reads
\be
L = \thalf\left(\partial_{0}\varphi \partial_{0}\varphi
- v^{2} \partial_{i} \varphi \partial_{i} \varphi 
- m^2 \varphi^{2} \right) +\ldots,
\ee
where the ellipsis stand for the self-interaction terms as well as
interactions with other fields. Notice that we introduced explicitly the
velocity parameter $v$. In microscopic studies of color
superconducting phases, which are valid at very large densities, 
this velocity is equal to $1/\sqrt{3}$ for all (pseudo-)  NG bosons. It is
smaller than $1$ because Lorentz symmetry is broken due to the finite value
of the quark chemical potential. By making use of the above Lagrangian
density, we derive the following expression for the heat current:
\be
u_{i} = \frac{\partial L}{\partial (\partial^{i} \varphi)}
\partial_{0} \varphi 
= v^{2} \partial_{i} \varphi \partial_{0} \varphi ,
\label{heat-cur}
\ee
This definition leads to the expression \cite{fgi} for the heat 
conductivity in terms of the corresponding correlator:
\be
\kappa_{ij} = -\frac{i}{2 T} \lim_{\Omega \to 0} \frac{1}{\Omega}
\left[\Pi_{ij}^{R}(\Omega+i\epsilon)-\Pi_{ij}^{A}(\Omega-i\epsilon)
\right],
\ee
where, in the Matsubara formalism, 
\ba
\Pi_{ij}(i\Omega_{m}) &=& v^{4} T \sum_{n} \int \frac{d^3 k}{(2\pi)^{3}}
k_{i} k_{j} i\Omega_{n} (i\Omega_{n} +i\Omega_{n-m}) \nonumber \\
&\times& S(i\Omega_{n}, \vec{k}) S(i\Omega_{n-m},\vec{k}) .
\ea
Here $\Omega_{n}\equiv 2\pi n T$ is the bosonic Matsubara frequency,
and $ S(i\Omega_{n}, \vec{k}) $ is the propagator of the (pseudo-) NG
boson. In general, the propagator should have the following form:
\be
S(\omega,\vec{k}) = \frac{1}
{(\omega+i\Gamma/2)^{2}-v^{2}\vec{k}^{2}-m^2},
\label{propagator}
\ee
where the width parameter $\Gamma(\omega, \vec{k})$ is related to the 
inverse lifetime (as well as the mean free path) of the boson.

In our calculation, it is very convenient to utilize the spectral
representation of the propagator,
\be
S(i\Omega_{n},\vec{k}) =\frac{1}{\pi} \int_{-\infty}^{\infty}
\frac{d\omega A(\omega,\vec{k})}{i\Omega_{n}-\omega}.
\ee
Then, the conductivity is expressed through the spectral function
$A(\omega,\vec{k})$ as follows:
\be
\kappa_{ij} = \frac{v^{4}}{2\pi T^2}  \int_{-\infty}^{\infty} 
\frac{\omega^2 d\omega }{\sinh^{2}\!\frac{\omega}{2T}}
\int \frac{d^3 k}{(2\pi)^{3}} k_{i} k_{j} 
A^{2}(\omega,\vec{k}).
\label{kap-general}
\ee
By making use of the explicit form of the propagator in
Eq.~(\ref{propagator}), we see that the spectral function of the 
(pseudo-) NG boson is
\ba
A(\omega,\vec{k}) &=& \frac{\omega\Gamma}{(\omega^{2}
- e_{k}^{2}-\Gamma^{2}/4)^{2}+\omega^{2}\Gamma^{2}} \nonumber\\
&\equiv & \frac{\Gamma}{4 e_{k}}
\left[\frac{1}{\left(\omega-e_{k}\right)^{2}
+\Gamma^{2}/4}-\left(\omega\to -\omega\right)\right],
\label{spectral-density}
\ea
where $e_{k} \equiv \sqrt{v^{2}k^2+m^{2}}$. By substituting this spectral
function into the general formula (\ref{kap-general}), we notice that the
off diagonal components of the thermal conductivity are zero and all the
diagonal components are equal. Both of these facts are a consequence of
the rotational symmetry of the system.  Thus the conductivity is
characterized by a single scalar quantity $\kappa$ which is introduced as
follows: $\kappa_{ij} =\kappa \delta_{ij}$. The explicit expression for
this scalar function reads
\be
\kappa = \frac{1}{48\sqrt{2} \pi^{2} v \Gamma T^2}  
\int_{0}^{\infty}\frac{\omega d\omega}
{\sinh^{2}\!\frac{\omega}{2T}} 
\left(\sqrt{X^{2}+\omega^{2}\Gamma^{2}}+X\right)^{3/2}\!,
\label{kap-final}
\ee
where we introduced the notation $X\equiv \omega^{2}-m^2-\Gamma^{2}/4$.
In the calculation leading to Eq.~(\ref{kap-final}), we made use of the 
result \cite{GrRyzh}:
\be
\int_{0}^{\infty} \frac{a^2 x^4 dx}
{\left[(x^2 + b)^{2}+a^2\right]^{2}}
=\frac{\pi}{8\sqrt{2}|a|}
\left(\sqrt{a^{2}+b^{2}}-b\right)^{3/2}.
\ee
For our purposes it will be sufficient to consider the conductivity
in the limit of a very small width. This is because the (pseudo-) NG
bosons in the CFL quark matter are weakly interacting. In this case
we derive the following approximate relation:
\be
\kappa = \frac{1}{24 \pi^{2} v T^2\Gamma} \int_{m}^{\infty}
\frac{d\omega \omega}{\sinh^{2}\!\frac{\omega}{2T}}
\left(\omega^{2}-m^2\right)^{3/2}.
\label{kap-G0}
\ee
It is interesting to notice that the same result can be easily derived 
from Eq.~(\ref{kap-general}). Indeed, in the limit of a very small width, 
{\it i.e.}, $\Gamma\to 0$, we can replace the square of the spectral 
function by a sum of two $\delta$-functions:
\be
A^{2}(\omega,\vec{k}) \to \frac{\pi}
{4\Gamma e_{k}^{2}} \left[
\delta\left(\omega-e_{k}\right)
+\delta\left(\omega+e_{k}\right)\right].
\ee
where the coefficients in front of the $\delta$-functions were 
determined unambiguously by using the following replacement rule: 
\be
\lim_{\Gamma\to 0}\frac{\Gamma^3}{(x^2+\Gamma^2/4)^2}
\to 4\pi\delta(x).
\ee
Now in the limit of small temperature, $T\ll m$, the result in 
Eq.~(\ref{kap-G0}) is given approximately by:
\ba
\kappa &=& \frac{m^{5}}{24\pi^{2} v \Gamma T^2} \int_{0}^{\infty}
\frac{x^{4} dx }{\sinh^{2}\!\frac{m\sqrt{1+x^2}}{2T}}\nonumber \\
&\simeq& \frac{m^{5/2}\sqrt{T}}
{2\sqrt{2}\pi^{3/2} v \Gamma} e^{-m/T}.
\label{kap-mass}
\ea
This demonstrates clearly that the contributions of heavy 
pseudo-NG bosons to the thermal conductivity are strongly suppressed. 
The largest contribution comes from the massless NG boson $\phi$ for 
which the thermal conductivity is
\be
\kappa_{\phi} = \frac{4 T^{3}}{3\pi^{2} v \Gamma_{\phi} } 
\int_{0}^{\infty} \frac{x^{4} dx }{\sinh^{2}\!x}
= \frac{2 \pi^{2}T^{3}}{45 v \Gamma_{\phi} }. 
\label{kap-m0}
\ee
In order to calculate $\kappa_{\phi}$ the width $\Gamma_{\phi}$ 
is required, or equivalently the mean free path $\ell_{\phi}$ since 
$\ell_{\phi} \equiv\bar{v}/\Gamma_{\phi}$, where 
$\bar{v}$ is the average thermal velocity of the particles responsible for 
heat transfer. This will be discussed in the next section.
For the moment we note that expression (\ref{kap-m0}) for the thermal 
conductivity is similar to the
phonon conductivity often used in solid state physics \cite{FrAn}.
It is also interesting to note that the result for the thermal
conductivity of both massive and massless bosons in Eqs.~(\ref{kap-mass})
and (\ref{kap-m0}) are consistent with a general relation from classical
Boltzmann kinetic theory \cite{Kittel}:
\be
\kappa = \oneth \bar{v}  c_{v} \ell,
\label{kap-Cv}
\ee
where $c_{v}$ is the specific heat.
In order to see that this relation holds, we note that the
average thermal velocity, 
$\langle\partial e_k/\partial k\rangle$, of massive bosons
is a factor $\sqrt{3T/m}$ smaller than the velocity of massless
bosons $v=1/\sqrt{3}$. After taking this
into account, as well as using the expressions for the specific heat in
Eqs.~(\ref{cV_bos_m}) and (\ref{cV_bos_0}) in the Appendix, we find that
the relation (\ref{kap-Cv}) between the conductivity and the specific heat 
is indeed satisfied. 

\section{Mean free path of the NG boson}
\label{mfp}

In the preceding section we derived a general expression for the boson 
portion of the thermal conductivity. The result involved the decay 
width of the (pseudo-) NG boson or equivalently the mean free path,
$\ell = \bar{v}/\Gamma$, for which we derive a simple estimate here. 
Using this we can understand the effect of the heat conductivity 
on the relaxation time of the temperature gradient inside the
quark core of a star. This is crucial for understanding the cooling
mechanism of compact stars with quark cores.

As we have remarked, the contribution of massive pseudo-NG
bosons to the thermal conductivity is suppressed by the exponential factor
$\exp(-m/T)$. In the CFL phase of quark matter, however, there is one
truly massless NG boson $\phi$ which should therefore give the dominant
contribution to the heat conductivity.
The interactions of $\phi$ with the CFL matter leads to a
finite value for its mean free path. Since this boson is a composite
particle there is always a non-zero probability at finite temperature
for its decay into a pair of
quark quasiparticles. It is natural to expect that such a process is
strongly suppressed at small temperatures, $T\ll\Delta$. This is confirmed
by a direct microscopic calculation in the region of asymptotic densities
which yields a decay width \cite{GS}:
\be
\Gamma_{\phi\to q q} (k) 
\simeq \frac{5\sqrt{2}\pi v k}{4(21-8\ln2)}
\exp\left(-\sqrt{\frac{3}{2}} 
\frac{\Delta}{T}\right) . 
\label{decay-quark}
\ee
If this were the only contribution, then the order of magnitude of
the mean free path of the NG boson would be 
\be
\ell_{\phi\to q q} \sim \frac{v}{T} \exp\left(\sqrt{\frac{3}{2}}
\frac{\Delta}{T}\right).
\ee
This grows exponentially with decreasing the temperature. For example, if
$T\lesssim \Delta/33$ the mean free path is $30$ km or more 
(in deriving this estimate we set $\Delta \simeq 50$ MeV). This 
scale is a few times larger than the typical size of a compact star. The
underlying physics in this calculation of the mean free path closely
resembles the propagation of sound waves in superfluid helium. We
recall that there are two types of excitations in superfluid helium:
gapless phonons and finite gap rotons. The main contribution to the
thermal conductivity comes from the phonons whose mean free path is mainly
determined by scattering on rotons \cite{Khalat}. The value of the mean
free path, as in the case of NG bosons, becomes exponentially large at
small temperatures ($T\ll \Delta$).

Now, in the case of the NG bosons the decay channel into quarks is not
the only contribution to the mean free path since 
the NG bosons can also scatter on one another. The amplitude has been 
derived by Son \cite{son} and is of order $k^4/\mu^{4}$
which gives a cross section of $\sigma_{\phi\phi} \simeq T^{6}/\mu^8$. 
This yields the following contribution to the width
of the NG boson:
\be
\Gamma_{\phi\phi} = v \sigma_{\phi\phi} n_{\phi}
\sim \frac{T^{9}}{\mu^8},
\label{self-int}
\ee
where $n_{\phi}$ is the equilibrium number density of the NG bosons. 
The explicit expression for $n_{\phi}$ is given in Eq.~(\ref{n_bos_0})
in the Appendix.
At small temperatures the scattering contribution in Eq.~(\ref{self-int})
dominates the width.
This leads to a mean free path 
\ba
\ell_{\phi\phi} \sim \frac{\mu^8}{T^{9}}
\approx 8\times 10^5\frac{\mu_{500}^8}{T_{MeV}^{9}}\mbox{ km}.
\ea
Here we defined the following dimensionless quantities:  $\mu_{500}
\equiv \mu/(500\mbox{ MeV})$, $\Delta_{50} \equiv \Delta/(50\mbox{ MeV})$,
and $T_{MeV} \equiv T/(1\mbox{ MeV})$. Both $\ell_{\phi\phi}$ and 
$\ell_{\phi\to qq}$ depend very strongly on temperature, however the 
salient point is that they are both larger than the size of a compact star
for temperatures $T_{MeV}$ of order 1.

We define $\tilde{T}$ to be the temperature at which the massive NG 
bosons decouple from the system. This is determined by 
the mass of the lightest pseudo-NG boson for which it is
not presently possible to give a reliable value.
Different model calculations \cite{SonSt,BS,SchMass} produce different
values which can range as low as 10 MeV. Thus, 
conservatively, we choose $\tilde{T}\simeq 1$ MeV.
Then the mean free path of the NG boson is comparable to 
or even larger than the size of a star for essentially all
temperatures $T\lesssim \tilde{T}$. It is also important 
to note that the mean free path is very sensitive to
temperature changes. In particular, at temperatures just a few times
higher than $\tilde{T}$ the value of $\ell$ may already become much
smaller than the star size. This suggests that, during the first few
seconds after the supernova explosion when the temperatures remain
considerably higher than $\tilde{T}$, a noticeable temperature gradient
may exist in the quark core. This should relax very quickly
because of the combined effect of cooling (which is very efficient at
$T\gg \tilde{T}$) and diffusion. After that almost the whole interior of
the star would become isothermal.

Before concluding this section, we point out that the geometrical
size of the quark phase limits the mean free path of the NG boson since
the scattering with the boundary should also be taken into
account. It is clear from simple
geometry that $\ell \sim R_{0}$, where $R_{0}$ is the radius of
the quark core. A similar situation is known to appear in high quality
crystals at very low temperatures \cite{Kittel}.

\section{Photon contributions}
\label{photon}

In this section, we discuss the role of photons in the CFL quark
core. It was argued in Ref.~\cite{JPS} that the mean free path of 
photons is larger than the typical size of a compact star at all
temperatures $T\lesssim \tilde{T}$. One might conclude therefore
that the photons would leave a finite region of the core in a very 
short amount of time after the core becomes transparent. 
If this were so, the photons would be 
able to contribute neither to thermodynamic nor to kinetic 
properties of the quark core. However
the neighboring neutron matter has very good metallic properties
due to the presence of a considerable number of electrons. As is known 
from plasma physics, low frequency electromagnetic waves cannot 
propagate inside a plasma. Moreover, an incoming electromagnetic 
wave is reflected from the surface of such a plasma \cite{Plasma}. 
In particular, 
if $\Omega_p$ is the value of the plasma frequency of the nuclear
matter, then all photons with frequencies $\omega < \Omega_p$ are 
reflected from the boundary. This effect is similar to the well 
known reflection of radio waves from the Earth's ionosphere.

The plasma frequency is known to be proportional to the square root 
of the density of charge carriers and inversely proportional to the
square root of their mass. It is clear therefore that the electrons,
rather than the more massive protons, will lead to 
the largest value of the plasma frequency in nuclear matter.
Our estimate for the value of this frequency is
\be
\Omega_p=\sqrt{\frac{4\pi e^2 Y_e \rho}{m_e m_p}}
\simeq 4.7\times 10^2 \sqrt{\frac{\rho Y_e}{\rho_{0}}}\mbox{ MeV}, 
\ee
where the electron density $n_e = Y_e \rho/m_p$ is given in 
terms of the nuclear matter density $\rho$ and the proton mass
$m_p$. Also $m_e$ denotes the electron mass, 
$Y_e\simeq 0.1$ is the number of electrons per baryon, 
and $\rho_{0}\approx 2.8 \times 10^{14} \mbox{ g cm}^{-3}$ 
is equilibrium nuclear matter density.

Since $\Omega_p$ is more than 100 MeV
electromagnetic waves of essentially all frequencies for temperatures
appropriate to the CFL phase will be reflected back into the core 
region. Thus photons present very early in the life of the star,
those produced by decays in the CFL matter as well as those 
produced at the nuclear interface will be trapped in the core.
In a way the boundary of the core looks like a good quality mirror
with some leakage which will allow a thermal photon distribution to
build up, even at the relatively high temperatures 
that existed during the first moments of stellar evolution.

Now, since photons are massless they also give a sizable contribution
to the thermal conductivity of the CFL phase. The corresponding 
contribution $\kappa_{\gamma}$ will be similar to the contribution of 
the massless NG boson in Eq.~(\ref{kap-m0}). Since the photons move
at approximately the speed of light \cite{Litim:2001mv} ($v\simeq1$ in 
our notation) and they have two polarization states, we obtain
\be
\kappa_{\gamma} =\frac{4 \pi^{2}T^{3}}{45 \Gamma_{\gamma}}.
\ee

Since the thermal conductivity is additive the total conductivity of
dense quark matter in the CFL phase is given by the  sum of the two 
contributions:
\be
\kappa_{CFL} =\kappa_{\phi} +\kappa_{\gamma}
\simeq \frac{2\pi^2}{9} T^{3} R_{0},
\ee
where for both a photon and a NG boson the mean free path
$\ell\sim R_0$.
This yields the value
\be
\kappa_{CFL} \simeq 1.2 \times 10^{32} T_{MeV}^{3} R_{0,km}
\mbox{ erg cm}^{-1} \mbox{sec}^{-1} \mbox{K}^{-1},
\ee
which is many orders of magnitude larger than the thermal conductivity
of regular nuclear matter in a neutron star \cite{van}.

\section{Stellar cooling}
\label{cool}

\begin{figure}
\begin{center}
\epsfxsize=8.0cm
\epsffile[135 65 330 265]{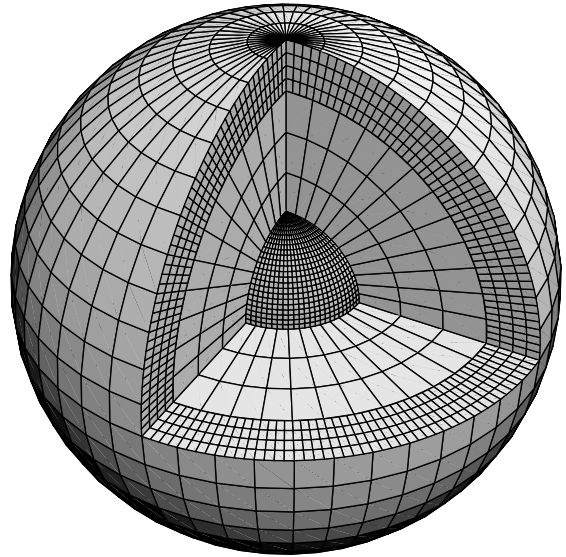}
\caption{The schematic structure of a compact star. At the center
is a quark core which is surrounded by ordinary  
nuclear matter. On the outside is a crust 
of nuclei, neutrons and electrons.}
\label{fig:1}
\end{center}
\end{figure}

In discussing the cooling mechanism for a compact star
we have to make some general assumptions about
the structure of the star. We accept without proof that
a quark core exists at the center of the star.  The radius of such a
core is denoted by $R_{0}$, while the radius of the whole star is 
denoted by $R$.
We exclude the possibility that the star is made
completely of CFL quark matter. Schematically, the internal structure
of the star is shown in Fig.~\ref{fig:1}. The quark core stays in direct
contact with the neighboring nuclear matter. From this nuclear contact
layer outwards the structure of the star should essentially be
the same as in an ordinary neutron star.
A detailed analysis of the interface between the quark core and the
nuclear matter was made in Ref.~\cite{Interface}. A similar analysis
might also be very useful for understanding the mechanism of heat
transfer from one phase to the other here.  For our purposes it is
sufficient to know only that direct contact between the quark and nuclear
phases is possible.  In fact the only assumption we need is that the
temperature is continuous across the interface.

Having spelled out the basic assumptions, we can consider
the physics that governs stellar cooling. Let us start from the moment
when the star is formed in a supernova explosion. Right after the
explosion a lot of high-energy neutrinos are trapped inside the star.
After about $10$ to $15$ seconds most of them escape from the star
by diffusion. The presence of the CFL quark core could slightly modify
the rate of such diffusion \cite{CarRed,Nu-diffusion,0203011}. By the end
of the deleptonization process, the temperature of the star would rise
to a few tens of MeV. Then, the star relatively quickly cools
down to about $\tilde{T}$ by the efficient process of neutrino emission. 
It is unlikely that the quark core would greatly affect the time scale for
this initial cooling stage. An ordinary neutron star would continue to
cool mostly by neutrino emission for quite a long time even after that
\cite{PLPSR}. Here we need to discuss how the presence of the CFL quark 
core affects the cooling process of the star after the temperature drops
below $\tilde{T}$.

Our result for the mean free path of the NG boson demonstrates that the
heat conductivity of dense quark matter in the CFL phase is very high. For
example, a temperature gradient of $1$ MeV across a core of $1$
km in size is washed away by heat conduction in a very short time
interval of order $R_{0,km}^2/v\ell(T) \simeq 6\times 10^{-4}$ sec. In
deriving this estimate, we took into account the fact that the specific
heat and the heat conductivity in the CFL phase are dominated by photons
and massless NG bosons and that $\ell\sim R_0$. In addition, we used the
relation in Eq.~(\ref{kap-Cv}). This estimate proves that, to a good
approximation, the quark core is isothermal at all temperatures $T\lesssim
\tilde{T}$.

The heat conductivity of the neighboring nuclear matter is also known to be
very high because of the large contribution from degenerate electrons
which have a very long mean free path. It is clear, then, that both the 
quark and the nuclear layers should be nearly isothermal with equal values 
of the temperature. When one of the layers cools down by any mechanism, the
temperature of the other will adjust almost immediately due to
the very efficient heat transfer on both sides of the
interface. This observation has a very important consequence for the
cooling rate of the star.
It will eventually be determined by the
combination of two effects: (i) neutrino emission from the nuclear part of
the star, and (ii) diffusion of photons through the outer crust made of
nuclei and non-degenerate electrons. Even if, for any reason the
intermediate nuclear layer happens to be abnormally thin photon diffusion 
will provide efficient cooling. The corresponding cooling time
would be of order $10^{7}$ years. This is many orders of magnitude shorter
than the estimated time of cooling of the CFL matter by neutrino
emission \cite{JPS}.

Now consider the order of magnitude of the cooling time for
a star with a CFL quark core. One of the most important components of
the calculation of the cooling time is the thermal energy of the star
which is the amount of energy that is lost in cooling. 
There are contributions to the total thermal energy
from both the quark and the nuclear parts of the star. The dominant
amount of thermal energy in the CFL quark matter is stored in photons
and massless NG bosons, see Eq.~(\ref{cV_bos_0}) and the brief discussion
in the Appendix. It is given by the following expression:
\ba
E_{CFL}(T) &=& \frac{4\pi R_{0}^{3}}{3}
\int_{0}^{T} c_v(T^{\prime}) dT^{\prime} \nonumber \\
&=&\frac{6(1+2v^3)T}{5} \left(\frac{\pi TR_{0}}{3v}\right)^{3},
\label{E_CFL}
\ea
where, for simplicity, we use the Newtonian approximation.
The specific heat here contains both $\phi$ and photon
contributions ($c_{v}=c_{v}^{(\phi)}+c_{v}^{(\gamma)}$)
which can be obtinaed from Eq. (\ref{cV_bos_0}) using the
appropriate degeneracies and velocities.
Numerically the expression in Eq.~(\ref{E_CFL}) leads to
the following result:
\be
E_{CFL}(T) \simeq 2.1 \times 10^{42} R_{0,km}^{3} T_{MeV}^{4} 
\mbox{ erg}.
\ee
Here $R_{0,km}$ is the quark core radius measured in kilometers. 
The thermal energy of the outer nuclear layer is provided mostly by 
degenerate neutrons. A numerical estimate is \cite{Shapiro}:
\be
E_{NM}(T) \simeq 8.1 \times 10^{49} \frac{M-M_{0}}{M_{\odot}}
\left(\frac{\rho_{0}}{\rho}\right)^{2/3} T^{2}_{MeV}
\mbox{ erg}, 
\ee
where $M$ is the mass of the star, $M_{0}$ is the mass of the quark
core and $M_{\odot}$ is the mass of the Sun. 
It is crucial to note that the thermal energy of the quark core is
negligible in comparison to that of the nuclear layer. 
Moreover, as the
star cools the ratio $E_{CFL}/E_{NM}$ will further decrease.

The second important component that determines stellar cooling is the
neutrino and/or photon luminosity which describes
the rate of energy loss. Typically, the neutrino luminosity dominates the
cooling of young stars when the temperatures are still higher than about
$10$ keV and after that the photon diffusion mechanism starts to dominate.
As argued in Ref.~\cite{JPS}, neutrino emission from the CFL quark
phase is strongly suppressed at low temperatures, {\it i.e.}, 
at $T\lesssim\tilde{T}$ in our notation. The neighboring 
nuclear layer, on the other hand, emits neutrinos quite efficiently. As 
a result, it cools relatively fast in the same way as an ordinary neutron 
star. The nuclear layer should be able to emit not
only its own thermal energy, but also that of the quark core which
constantly arrives by the very efficient heat transfer process. The 
analysis of this cooling mechanism, however, is greatly simplified by the 
fact that the thermal energy of the quark core is negligible compared to 
the energy stored in the nuclear matter.
By making use of the natural assumption that local neutrino emissivities
from the nuclear matter are not affected by the presence of the quark core,
we conclude that the cooling time of a star with a quark core
is essentially the same as for an ordinary neutron star provided 
that the nuclear layer is not extremely thin.

\section{Conclusions}
\label{conclusion}

As is clear from our analysis, the thermal conductivity of the CFL color
superconducting dense quark matter is very high for typical values of the
temperature found in a newborn compact star. This is a direct consequence
of the existence of the photon and the massless NG boson associated with 
the breaking of baryon number. (Note that the quark contribution to the 
thermal conductivity is strongly suppressed in the CFL phase, as discussed 
in Ref.~\cite{BGV}.) The mean free path of the NG boson appears to be 
very large as a result of its weak self-interaction and its weak
interaction with the rest of the matter in the CFL phase. The same is 
true for the photon. Our rough
estimate shows that even at a relatively early stage of 
stellar cooling when the temperature is of order $\tilde{T}$, say 1 MeV or 
so, the mean free paths are already larger than the size of the star. It
should be noted, however, that the NG bosons cannot leave the CFL phase
and escape from the core of the star carrying away their energy. By their
nature, these NG bosons are collective excitations of the CFL phase and
thus they are ``confined" to the medium which made their existence
possible. In this respect, they resemble phonons in solid state physics
which also, by their nature, cannot escape from inside a crystal. 
Although for a different reason, thermal photons also cannot leave 
the finite region of the quark core. They are efficiently 
reflected from the electron plasma of the neighboring nuclear matter. 

We mention that the (pseudo-) NG bosons and photons should also dominate
other kinetic properties of dense quark matter in the CFL phase. For 
example, the shear viscosity should be mostly due to photons and the 
same massless NG bosons associated with the breaking of baryon number.
The electrical conductivity, on the other hand, would be mostly 
due to the lightest {\em charged} pseudo-NG boson, {\it i.e.}, 
the $K^{+}$. Thus, in the limit of small temperatures $T\to 0$ 
the electrical conductivity will be suppressed by a factor 
$\exp(-m_{K^{+}}/T)$.

Since the neutrino emissivity of the CFL core is strongly
suppressed, the heat is transferred to the outer nuclear layer
only through the direct surface contact. While both the core and the
outer layer contribute to the heat capacity of the star, it is only
the outer layer which is capable of emitting this heat energy
efficiently in the form of neutrinos. The combination of these two factors
tends to extend the cooling time of a star. However, because of the very 
small thermal energy of the quark core compared to that of the
nuclear matter, the time scale for cooling would only get noticeably shorter
than that for an ordinary neutron star when the quark core radius is
nearly the same as the stellar radius. Thus it appears that the cooling 
of stars with CFL quark cores will differ little from the cooling of 
standard neutron stars. A similar conclusion has been reached for stars 
with regular, non-CFL quark interiors \cite{ppls}.

In passing it is interesting to speculate about the possibility that a bare 
CFL quark star made entirely of dense quark matter could exist.
If it were possible, it would look like a transparent dielectric
\cite{RW-diamond}. Our present study suggests such a star would also
have very unusual thermal properties. Indeed, if the star has a finite
temperature $T\lesssim \tilde{T}$ after it was created, almost all of its
thermal energy would be stored in the NG bosons. Notice that all the 
photons would leave the star very soon after transparency set
in because the star is assumed to have no nuclear matter 
layer. The local interaction as 
well as the self-interaction of the NG bosons is very weak so that 
we argued in Sec.~\ref{mfp} that their mean free path would be limited only 
by the geometrical size of the star. This suggests that the only potential 
source of energy loss in the bare CFL star would be the interaction of the NG
bosons at the star boundary, together with very strongly suppressed 
\cite{JPS} photon and neutrino emission. 
It is likely, therefore, that such stars might be very dim and 
might even be good candidates for some of the dark matter in the 
Universe. A more detailed discussion of this issue is, however, beyond the 
scope of the present paper.

Finally, we would also like to note that the thermal conductivity of
the 2SC color superconducting phase of two lightest quarks should also 
be very large, but for a different reason. In this case there are quarks
which do not participate in color condensation and thus give rise to
gapless quasiparticles. The corresponding contribution to the heat
conductivity (which is similar to the contribution of degenerate electrons
inside nuclear matter) should be very large. Therefore, if it were
possible for the core of a compact star to be made of the 2SC quark phase,
it would again be nearly isothermal. The results of Ref.~\cite{alf}
suggest, however, that the 2SC phase cannot appear in a compact star.

\section*{Acknowledgments}
We would like to thank Yong Qian for many interesting 
discussions and M. Prakash, S. Reddy and D. T. Son for useful comments 
on the manuscript.  This work was supported by the U.S. Department of
Energy Grant No.~DE-FG02-87ER40328.

\appendix 

\section{Thermodynamical quantities}

In this Appendix, for convenience we list 
expressions for the thermodynamic quantities that we use throughout 
the main text of the paper.

It is most convenient to start with the expression for the pressure.
The other quantities (entropy, free energy, {\it etc.}) can then be 
easily derived \cite{Kapusta}. In the case of a massive boson field,
the pressure is
\be
P^{(bos)}=- T \int\frac{d^3 k}{(2\pi)^{3}}\ln(1-e^{-e_{k}/T}),
\label{P_bos}
\ee
where $e_{k}=\sqrt{v^2 k^2+m^2}$. The corresponding energy 
density, number density and the specific heat are:
\ba
U^{(bos)}&=&\int\frac{d^3 k}{(2\pi)^{3}} \frac{e_{k}}{e^{e_{k}/T}-1} , 
\label{U_bos} \\
n^{(bos)}&=& \int\frac{d^3 k}{(2\pi)^{3}}\frac{1}{e^{e_{k}/T}-1},
\label{n_bos} \\
c_{v}^{(bos)}&=& \frac{1}{4T^2} \int\frac{d^3 k}{(2\pi)^{3}}
\frac{e_{k}^{2}}{\sinh^{2}\!\frac{e_{k}}{2T}} .
\label{cV_bos} 
\ea
In the case of a small temperature (as compared to the value of mass), 
these thermodynamic quantities approach the following asymptotic 
expressions \cite{thermo-asympt}:
\ba
P^{(bos)} 
&\simeq & \frac{m^{3/2}T^{5/2}}{2\sqrt{2}\pi^{3/2}v^{3}} e^{-m/T},
\label{P_bos_m} \\
U^{(bos)} 
&\simeq &\frac{m^{5/2}T^{3/2}}{2\sqrt{2}\pi^{3/2}v^{3}} e^{-m/T}, 
\label{U_bos_m} \\
n^{(bos)}
&\simeq &\frac{m^{3/2}T^{3/2}}{2\sqrt{2}\pi^{3/2}v^{3}}e^{-m/T} , 
\label{n_bos_m} \\
c_{v}^{(bos)} 
&\simeq & \frac{m^{7/2}}{2\sqrt{2}\pi^{3/2}v^{3}\sqrt{T}}e^{-m/T}.
\label{cV_bos_m} 
\ea
And, in the limit of massless particles, they are \cite{thermo-asympt}
\ba
P^{(bos)} &=&
\frac{\pi^2 T^4}{90 v^3} ,
\label{P_bos_0} \\
U^{(bos)} &=& 
\frac{\pi^2 T^4}{30v^3} ,
\label{U_bos_0} \\
n^{(bos)} &=& 
\frac{\zeta(3) T^{3}}{\pi^{2}v^{3}},
\label{n_bos_0} \\
c_{v}^{(bos)} &=& 
\frac{2\pi^{2} T^{3}}{15v^{3}},
\label{cV_bos_0} 
\ea
where the value of the Riemann function $\zeta(3) \approx 1.202$.
For completeness, let us also present some thermodynamical properties 
of color superconducting quarks. In this case, it is very convenient to 
separate the zero and finite temperature contributions explicitly. 
For the pressure of a single quark with a gap $\Delta$, we find: 
\ba
P_{qrk} &=& P^{(0)}_{qrk} 
+ 2T \int\frac{d^3 k}{(2\pi)^{3}}\ln(1+e^{-E^{(-)}_{k}/T})
\nonumber \\
&+& 2 T \int\frac{d^3 k}{(2\pi)^{3}}\ln(1+e^{-E^{(+)}_{k}/T}),
\label{P_qrk}
\ea
where $E^{(\pm)}_{k} =\sqrt{(e_{k}\pm \mu)^2+\Delta^2}$ are the 
quasiparticle energies and $\mu$ is the chemical potential. 
The zero temperature contribution is approximately given by
\begin{equation}
P^{(0)}_{qrk} = \int\frac{d^3 k}{(2\pi)^{3}}
\left(E^{(+)}_{k}+E^{(-)}_{k}\right)-\alpha\Delta^2\;.
\label{pot}
\end{equation}
The last term is needed so that the gap equation is obtained when 
the pressure is maximized with respect to variations in 
$\Delta$. Using the gap equation to eliminate 
$\alpha$ and renormalizing by removing the divergent term, 
the pressure becomes
\begin{equation}
P^{(0)}_{qrk} \simeq \frac{\mu^{4}}{12\pi^{2}} 
+ \frac{\mu^{2} \Delta^2}{4\pi^2} +\ldots\;,
\label{P_qrk_0}
\end{equation}
in agreement with the results obtained using the framework of
the Nambu-Jona-Lasinio model \cite{RW-diamond} and directly
from quantum chromodynamics \cite{mira}.
The ellipsis denotes subleading terms suppressed by powers 
of either $(m/\mu)^2$ or $(\Delta/\mu)^2$.
As is easy to check, at small
temperatures, i.e., $T\ll \Delta$, the finite temperature corrections 
in Eq.~(\ref{P_qrk}) are suppressed by a factor 
$\exp(-\Delta/T)$. Thus, the above 
zero temperature contribution will dominate the pressure of the 
CFL phase. Similarly, the main contribution to the number density
and the energy density of CFL matter at $T\ll \Delta$ comes from the
zero temperature terms. Using Eq. (\ref{P_qrk_0}) the $T=0$ contribution
to the number density is
\be
n^{(0)}_{qrk} \simeq \frac{\mu^3}{3\pi^2}+\frac{\mu \Delta^2}{2\pi^2}
\left(1+\frac{\mu}{\Delta}\frac{\partial\Delta}{\partial\mu}
\right)\;.
\ee
Note that the partial derivative term is needed since the gap equation,
which specifies $\Delta(\mu)$, was used in arriving at 
Eq. (\ref{P_qrk_0}). This term contributes at the same order (unity)
as the first term in parentheses. The energy density is then
\be
U^{(0)}_{qrk} \simeq \frac{\mu^{4}}{4\pi^2}+\frac{\mu^2\Delta^2}{4\pi^2}
\left(1+\frac{2\mu}{\Delta}\frac{\partial\Delta}{\partial\mu}\right)\;.
\ee
Here it is appropriate to note that all thermodynamic
functions which are defined through the temperature derivatives of 
either the pressure ({\it e.g.}, entropy) or the energy ({\it e.g.}, 
specific heat) are exponentially suppressed at $T\ll \Delta$. Therefore, 
the dominant contributions to this second type of functions would come
from (pseudo-) NG bosons in the CFL phase of dense quark matter.


\begin{references}

\item[*]{On leave of absence from Bogolyubov Institute for
Theoretical Physics, 252143, Kiev, Ukraine.}

\bibitem{ColPer} J.C.~Collins and M.J.~Perry,
Phys.\ Rev.\ Lett.\  {\bf 34}, 1353 (1975).

\bibitem{old} B.~C.~Barrois, 
Nucl.\ Phys.\ B {\bf 129}, 390 (1977);
S.~C.~Frautschi,
in ``Hadronic matter at extreme energy density", edited by 
N.~Cabibbo and L.~Sertorio (Plenum Press, 1980);
D.~Bailin and A.~Love,
Phys.\ Rept.\  {\bf 107}, 325 (1984).

\bibitem{W1S1} M.~G.~Alford, K.~Rajagopal and F.~Wilczek,
Phys.\ Lett.\ B {\bf 422}, 247 (1998);
R.~Rapp, T.~Sch\"{a}fer, E.~V.~Shuryak and M.~Velkovsky,
Phys.\ Rev.\ Lett.\  {\bf 81}, 53 (1998).
 
\bibitem{PR1-Son} D.~T.~Son,
Phys.\ Rev.\ D {\bf 59}, 094019 (1999);
R.~D.~Pisarski and D.~H.~Rischke,
Phys.\ Rev.\ Lett.\  {\bf 83}, 37 (1999).

\bibitem{2nd-wave} T.~Schafer and F.~Wilczek,
Phys.\ Rev.\ D {\bf 60}, 114033 (1999);
D.~K.~Hong, V.~A.~Miransky, I.~A.~Shovkovy and L.~C.~R.~Wijewardhana,
Phys.\ Rev.\ D {\bf 61}, 056001 (2000);
erratum {\em ibid.}\ D {\bf 62}, 059903 (2000);
R.~D.~Pisarski and D.~H.~Rischke,
Phys.\ Rev.\ D {\bf 61}, 051501 (2000).

\bibitem{3rd-wave} S.~D.~Hsu and M.~Schwetz,
Nucl.\ Phys.\ B {\bf 572}, 211 (2000);
W.~E.~Brown, J.~T.~Liu and H.~C.~Ren,
Phys.\ Rev.\ D {\bf 61}, 114012 (2000).

\bibitem{us2} I.~A.~Shovkovy and L.~C.~R.~Wijewardhana,
Phys.\ Lett.\ B {\bf 470}, 189 (1999);
T.~Sch\"{a}fer,
Nucl.\ Phys.\ B {\bf 575}, 269 (2000).

\bibitem{RW-Al-rev} K.~Rajagopal and F.~Wilczek,
arXiv:hep-ph/0011333;
M.~G.~Alford,
Ann.\ Rev.\ Nucl.\ Part.\ Sci.\  {\bf 51}, 131 (2001).

\bibitem{alf} M.~Alford and K.~Rajagopal,
arXiv:hep-ph/0204001.

\bibitem{ARW} M.~Alford, K.~Rajagopal and F.~Wilczek,
Nucl.\ Phys.\ B {\bf 537}, 443 (1999).

\bibitem{CasGat} R.~Casalbuoni and R.~Gatto,
Phys.\ Lett.\ B {\bf 464}, 111 (1999).

\bibitem{SonSt} D.~T.~Son and M.~A.~Stephanov,
Phys.\ Rev.\ D {\bf 61}, 074012 (2000);
erratum {\em ibid.} D {\bf 62}, 059902 (2000).

\bibitem{BS} P.~F.~Bedaque and T.~Sch\"{a}fer,
Nucl.\ Phys.\ A {\bf 697}, 802 (2002);
D. B. Kaplan and S. Reddy, Phys. Rev. D {\bf65}, 054042 (2002).

\bibitem{MirSho} V.~A.~Miransky and I.~A.~Shovkovy,
Phys.\ Rev.\ Lett.\ {\bf 88}, 111601 (2002) [arXiv:hep-ph/0108178];
T.~Schafer, D.~T.~Son, M.~A.~Stephanov, D.~Toublan and J.~J.~Verbaarschot,
Phys.\ Lett.\ B {\bf 522}, 67 (2001) [arXiv:hep-ph/0108210].

\bibitem{Son-s} D.~T.~Son,
arXiv:hep-ph/0108260.

\bibitem{JPS} P.~Jaikumar, M.~Prakash and T.~Sch\"{a}fer,
arXiv:astro-ph/0203088.

\bibitem{fgi} E. J. Ferrer, V. P. Gusynin and V. de la Incera,
arXiv: cond-matt/0203217.

\bibitem{GrRyzh} I.~S.~Gradshteyn and I.~M.~Ryzhik, 
{\sl Tables of Integrals, Series and Products} (Academic, 
New York, 1965) 3.252.9.

\bibitem{FrAn} J.~J.~Freeman and A.~C.~Anderson,
Phys. Rev. B {\bf 34}, 5684 (1986).

\bibitem{Kittel} C.~Kittel, {\sl Introduction to solid state physics},
(John Wiley \& Sons, Inc., 1960) p. 139.

\bibitem{GS} V.~P.~Gusynin and I.~A.~Shovkovy,
Nucl.\ Phys.\ A {\bf 700}, 577 (2002).

\bibitem{Khalat} I. M. Khalatnikov, {\sl An introduction 
to the theory of superfluidity}, (Addison-Wesley Pub. Co., 1989).

\bibitem{son} D. T. Son, hep-ph/0204199.

\bibitem{SchMass} T.~Schafer,
arXiv:hep-ph/0201189.

\bibitem{Plasma} P.~A.~Sturrock, {\sl Plasma Physics}, 
(Cambridge University Press, 1994).

\bibitem{Litim:2001mv} D.~F.~Litim and C.~Manuel,
Phys.\ Rev.\ D {\bf 64}, 094013 (2001).

\bibitem{van} J. M. Lattimer, K. A. Van Riper, M. Prakash and M. Prakash,
Astrophys.\ J.\  {\bf425}, 802 (1994).
 
\bibitem{Interface} M.~G.~Alford, K.~Rajagopal, S.~Reddy and F.~Wilczek,
Phys.\ Rev.\ D {\bf 64}, 074017 (2001).

\bibitem{CarRed}
G.~W.~Carter and S.~Reddy,
Phys.\ Rev.\ D {\bf 62}, 103002 (2000).

\bibitem{Nu-diffusion} A.~W.~Steiner, M.~Prakash and J.~M.~Lattimer,
Phys.\ Lett.\ B {\bf 509}, 10 (2001)
[arXiv:astro-ph/0101566].

\bibitem{0203011} S.~Reddy, M.~Sadzikowski and M.~Tachibana,
arXiv:nucl-th/0203011.

\bibitem{PLPSR} 
M.~Prakash, J.~M.~Lattimer, J.~A.~Pons, A.~W.~Steiner and S.~Reddy,
Lect.\ Notes Phys.\  {\bf 578}, 364 (2001).

\bibitem{Shapiro} S.~L.~Shapiro and S.~A.~Teukolsky, {\sl Black
holes, white dwarfs, and neutron stars: the physics of compact
objects}, (John Wiley \& Sons, 1983).

\bibitem{BGV} D.~Blaschke, H.~Grigorian and D.~N.~Voskresensky,
Astron.\ Astrophys.\  {\bf 368}, 561 (2001).

\bibitem{ppls} D. Page, M. Prakash, J. M. Lattimer and A. W. Steiner,
Phys. Rev. Lett {\bf85}, 2048 (2000).

\bibitem{RW-diamond} K.~Rajagopal and F.~Wilczek,
Phys.\ Rev.\ Lett.\  {\bf 86}, 3492 (2001).

\bibitem{Kapusta} J.~I.~Kapusta, {\sl Finite-temperature field 
theory}, (Cambridge University Press, 1989).

\bibitem{thermo-asympt} S.~M.~Johns, P.~J.~Ellis and J.~M.~Lattimer,
Astrophys.\ J.\  {\bf 473}, 1020 (1996).

\bibitem{mira} From the effective potential for four quarks at the 
minimum in V. A. Miransky, I. A. Shovkovy and L. C. R. 
Wijewardhana, Phys. Lett. B {\bf468}, 270 (1999).
\end{references}
\end{document}